\documentclass[12pt]{article}
\usepackage[reqno]{amsmath}
\usepackage{amsthm}
\usepackage[dvips]{color,graphicx}
\usepackage{amssymb} 
\usepackage{epsfig}
\def\href#1#2{#2}
\def\IP{\relax{\rm I\kern-.18em P}}

\def\tB{{\rm B}}

\def\im{\tilde{\imath}}
\newlength{\textlength}

\newcommand{\myframebox}[3]{\fbox{\parbox[c][#1][c]{#2}{\makebox[#2]{#3}}}}

\newcommand{\LA}[1]{{\mathfrak{#1}}}

\def\End{{\rm End}}
\def\Hom{{\rm Hom}}  

\def\bq{{\bar q}}
\def\Fun{{\it Fun}}  

\def\cS{{\sf S}}
\def\tF{{\rm F}}
\def\tf{{\rm f}}
\def\tCS{{\rm CS}}
\def\tA{{\rm A}}
\def\tL{{\rm L}}

\def\astk{\, ,\, }

\def\cE{{\cal E}}

\def\bD{{\overline{D}}}
\def\sh{{\LA{h}}}
\def\d{\delta} 
\def\hh{{\widehat{\sh}}}
\def\bL{{\bar L}}

\newcommand{\beq}{\begin{equation}}
\newcommand{\eeq}{\end{equation}}
\newcommand{\beqa}{\begin{eqnarray}}
\newcommand{\eeqa}{\end{eqnarray}}
\def\La{L}
\def\la{l} 
\def\bz{{\bar z}}
\def\cJ{{\cal J}}
\def\H{{\cal H}}

\def\sg{{\LA{g}}} 

\def\hg{{\widehat{\sg}}} 

\def\bz{{\bar z}}

\def\dim{{\rm dim\,}}

\def\c{\gamma}

\def\cH{{\cal H}}
\def\cJ{{\cal J}}

\newcommand{\nn}{\nonumber}

\def\a{\alpha}
\def\b{\beta} 
\def\IP{\relax{\rm I\kern-.18em P}}
\def\tr{{\rm tr\ }}

\def\QC{\mathbb{C}}

\def\QZ{\mathbb{Z}}

\def\nn{\nonumber}

\def\cH{{\cal H}}

\def\a{\alpha}

\input epsf
\newcount\figno
\def\fig#1#2#3{
\par\begingroup\parindent=0pt\leftskip=1cm\rightskip=1cm\parindent=0pt
\baselineskip=11pt
\global\advance\figno by 1
\epsfxsize=#3
\centerline{\epsfbox{#2}}
\vskip 12pt
{\bf Figure \the\figno:} #1\par
\endgroup\par
}
\def\figlabel#1{\xdef#1{\the\figno 
\mbox{ }}}
\def\encadremath#1{\vbox{\hrule\hbox{\vrule\kern8pt\vbox{\kern8pt
\hbox{$\displaystyle #1$}\kern8pt}
\kern8pt\vrule}\hrule}}
\newcommand{\email}[1]{{\footnote{email: #1}}}

\begin{document} 
\title{\bf D-Branes in Coset Models}
\author {{\sc Stefan Fredenhagen}\\ \small  Max-Planck-Institut f{\"u}r 
              Gravitationsphysik, Albert-Einstein-Institut
      \\ \small Am M{\"u}hlenberg 1, D--14476 Golm, Germany      
        \email{stefan@aei.mpg.de} \\[4mm] 
        {\sc Volker Schomerus}\\ 
 \small Laboratoire de Physique Th{\'e}orique de l'{\'E}cole Normale 
  Sup{\'e}rieure \\ \small
 24 rue Lhomond, F-75231 Paris Cedex 05, France \  \& \\
 \small Max-Planck-Institut f{\"u}r Gravitationsphysik, 
 Albert-Einstein-Institut\\ \small 
 Am M{\"u}hlenberg 1, D-14476 Golm, Germany   
 \email{vschomer@aei.mpg.de}} 
\date{November 22, 2001} 


\maketitle
\addtolength{\baselineskip}{1pt} 
\begin{abstract}
The analysis of D-branes in coset models $G/H$ provides 
a natural extension of recent studies on branes in WZW-theory and it 
has various interesting applications to physically relevant models. 
In this work we develop a reduction procedure that allows to 
construct the non-commutative gauge theories which govern the 
dynamics of branes in $G/H$. We obtain a large class of solutions
and interprete the associated condensation processes geometrically. 
The latter are used to propose conservation laws for the dynamics
of branes in coset models at large level $k$. In super-symmetric
theories, conserved charges are argued to take their values in the 
representation ring of the denominator theory. Finally, we apply 
the general results to study boundary fixed points in two examples, 
namely for parafermions and minimal models. \vspace*{-19cm} 
\end{abstract} 
AEI-2001-136 \hfill LPTENS 01/43 \\ hep-th/0111189
\newpage

\addtolength{\baselineskip}{4pt} 
\section{Introduction}
The study of D-branes on compact group manifolds has taught us a 
lot about the classification and the dynamics of branes in curved 
but highly symmetric backgrounds (see e.g.\ \cite{KliSev,AlSc1,Gaw1,
Sta2,AlReSc1,FFFS1,AlReSc2,BaDoSc,Paw}). Part of this analysis applies 
directly to relevant string backgrounds, namely to string theory in 
near horizon geometries of NS5 or D3 branes. In both cases, a 
3-sphere, i.e.\ the group manifold of SU(2), appears as part 
of the background. 
\smallskip

But there is a more important aspect of these developments 
that is deeply rooted in CFT model building. In fact, the 
WZW models that are used to describe strings on group 
manifolds appear as the basic building blocks for all the 
coset and orbifold constructions of exactly solvable string 
backgrounds. It is therefore natural to analyze how much of 
the known properties of strings and branes on group manifolds 
descends down to less symmetric coset spaces $G/H$. This has 
been initiated in a number of recent papers \cite{Sta1,FrSc2,
MaMoSe1,Gaw2,EliSar} and it is our main subject below.    
\smallskip

Brane configurations (of so-called ``Cardy type'') in a coset 
model $G/H$ can be labeled by representations $P$ of the Lie 
algebra $\sg \oplus \sh$. Let us remark that not all 
representations are to be admitted here and that there 
exist some identifications between representations that 
are associated with exactly the same brane configuration. 
These issues will be addressed in more detail in Section 3. 
Admissible irreducible representations $P$ of $\sg \oplus 
\sh$ correspond to elementary branes while reducible
representations enter when we want to describe 
superpositions of the elementary branes. 
\smallskip

In Section 4 we shall construct the effective non-commutative
gauge theory for such branes $P$ in a certain limiting regime 
of the coset model. These theories are obtained by some kind 
of dimensional reduction from the fuzzy gauge theories that 
control the dynamics of Cardy type branes on group manifolds 
\cite{AlReSc2,AlReScrev}. They are (constrained) matrix models 
involving a Yang-Mills and a Chern-Simons like term.
\smallskip 

A large number of solutions to these non-commutative gauge 
theories is constructed and interpreted as a formation of 
bound states in Section 5. We shall show that two brane
configurations $P$ and $Q$ are related by condensation if the 
representations one obtains by restriction to the diagonally 
embedded $\sh \subset \sg \oplus \sh$ are equivalent, 
$$ P |_\sh \ \cong \ Q |_\sh  \ \ \ \ \ 
   \mbox{ where } \ \ \ \sh = \sh_{\rm diag} \subset 
   \sg \oplus \sh \ \ \ . $$
Here, the subscript `diag' refers to the diagonal embedding 
$X \mapsto X \oplus X$ of $\sh$ into $\sg \oplus \sh$. For the 
criterion to make sense, we use the identification between brane 
configurations and representations of $\sg \oplus \sh$. 
\smallskip

In non-supersymmetric backgrounds there will be many other 
processes that involve tachyonic (relevant) fields and these 
lead to further relations between brane configurations. 
But in super-symmetric models there is a fairly good chance 
that the criterion above exhausts all the possibilities. In 
this case, our findings suggest that the conserved charges of 
Cardy-type D-branes in a coset $G/H$ take their values in 
the representation ring of the denominator $H$, i.e.\ there 
is one conserved charge for each irreducible representation 
of $H$. 
This also fits nicely with the structure of Ramond-Ramond charge lattices
found in certain Kazama-Suzuki models \cite{LerWal}.
For a trivial denominator $H = {e}$, there appears 
just one irreducible representation and hence we recover 
the known result that Cardy type branes on group manifolds
carry only D0-brane charge. At present, our studies of the 
charge group in coset models are restricted to a limiting 
regime and one expects that they receive the same type of 
corrections that appear for group manifolds 
\cite{AlSc2,Sta3,Sta4,FrSc1,MaMoSe2}.   
\smallskip 

In the last section of this paper we shall apply our general 
results to branes in parafermion and minimal models. For 
minimal models our general results will provide a large 
number of new candidates for flows between boundary theories
extending previous related work in conformal field theory
\cite{LeSaSi,DRTW,ReRoSc,GrRuWa1,GrRuWa2} (see also 
\cite{Graham}) and we shall propose a nice and very suggestive 
geometrical interpretation. The results on parafermions and N=2 
super-conformal minimal models have been (partly) announced 
before in \cite{FrSc2}.

\section{Review of branes in group manifolds} 
\setcounter{equation}{0} 
\label{sec:review}
In this section we shall review some of the results on branes
in a group manifold $G$. Strings moving on $G$ are described 
by the WZW-model and so we will start by recalling some facts 
about the affine Lie algebras obtained from chiral currents in 
these theories. Following the work of Cardy \cite{Car}, we 
shall then present the solution of the boundary WZW problem
and its geometric interpretation \cite{AlSc1}. Finally, we 
discuss the low energy effective action of branes on group 
manifolds \cite{AlReSc2}.   
\smallskip

WZW models for a compact simple simply connected group $G$ are 
parametrized by one discrete parameter $k$ which is known as 
the {\em level}. We can think of $k$ as controlling the volume 
(or `size') of the group manifolds. From the basic group valued 
field $g$ of the model one can construct (anti-)holomorphic 
currents $J, \bar J$ taking values in the Lie algebra $\sg$. 
Throughout this work we are interested in boundary theories in 
which the currents are subjected to the following boundary 
condition at $z = \bz$  
\begin{equation} J^\a(z) \ = \ \bar J^\a(\bz) \ \ \mbox{ for } \ \ 
   \a = 1,\dots, D = \dim G \ \ . 
\label{eq:glue} \end{equation}
These boundary conditions were shown in \cite{AlSc1} to 
describe branes localized along conjugacy classes of the 
group and they are equipped with a non-vanishing B-field. 
The stability of the associated super-symmetric theories 
was established in \cite{BaDoSc,Paw}, at least in the 
large volume regime. At finite level $k$, they can be 
shown to possess a tachyon free spectrum 
\cite{AlReSc1,AlReSc2}. 
\smallskip

The WZW-model with boundary condition (\ref{eq:glue}) can be 
solved using ideas that go back to the work of Cardy
\cite{Car} and Runkel \cite{Run1} (see \cite{AlReSc1,FFFS1} 
for applications to the WZW model). The solution uses data 
from the representation theory of affine Lie algebras which 
we shall recall briefly also to set up our notations. We 
shall label sectors of the theory by elements $l$ taken 
from a finite set $\cJ^\sg_k$. The corresponding state 
spaces $\cH^\sg_\la$ are generated from irreducible 
representations $V^\la \subset \cH^\sg_\la$ of the finite 
dimensional Lie algebra $\sg = {\rm Lie} G $. This implies 
that the sectors of the theory at finite $k$ form a subset 
of the set $\cJ^\sg$ of irreducible representations of $\sg$, 
i.e.\ $ \cJ^\sg_k \subset \cJ^\sg$. We will often identify 
the elements $\la \in \cJ^\sg_k$ with the corresponding 
element of $\cJ^\sg$. 

On $\cH^\sg_\la$ there exists an action of the Virasoro 
algebra whose generators we denote by $L^\sg_n$ making 
explicit reference to the Lie algebra $\sg$. The space $V^\la 
\subset \cH^\sg_\la$ consists of ground states with conformal 
dimension $h^\sg_\la \sim \Delta_\la / (k+c^\vee)$ where 
$\Delta_\la$ is the value of the quadratic Casimir in the
representation $\la \in \cJ^\sg$ and $c^\vee$ denotes the dual 
Coxeter number. As we send $k$ to infinity, the conformal
dimension of these ground states vanishes. 
\medskip

Let us now consider the WZW-model associated with the 
diagonal modular invariant, 
$$ 
  Z(q,\bq) \ = \ \sum_{\la \in \cJ^\sg} \ |\chi^\sg_\la(q)|^2\ \ ,  
$$  
where $\chi^\sg_\la(q)$ denotes the character of the sector 
$\la$. The choice of $Z$ implies that the bulk fields of the 
boundary theories we are about to discuss are obtained as 
descendants of primary fields $\Phi^{\la,\la}(z,\bz) = 
\Phi^\la(z,\bz)$, one for each sector of the affine Lie 
algebra. The space of bulk fields comes equipped with two 
commuting representations of the affine Lie algebra. Below, 
we shall frequently make use of the descendants 
\begin{equation} \Phi^\la_{nm} (z,\bz)  \ \ \mbox{ for } \ \ \la \in 
   \cJ_k 
\label{eq:list1} \end{equation}
and $n,m$ each label vectors from a basis of the representation 
space $V^\la$. The fields in the list (\ref{eq:list1}) are obtained 
from the primary fields by acting with zero modes of the two 
commuting affine Lie algebras. They correspond to ground states 
in the bulk theory and their conformal dimensions are $(h,\bar h) 
= (h^\sg_\la,h^\sg_\la)$.  
\medskip

Following the analysis of Cardy, the boundary WZW model
with condition (\ref{eq:glue}) admits as many different 
solutions as there are sectors $\la \in \cJ^\sg_k$. We 
will denote the boundary theories by capital letters
$\La, \dots$. These boundary theories can be 
characterized by the 1-point functions of bulk fields, 
i.e.\ by the the coupling of closed string modes to the 
brane. According to \cite{Car}, these couplings are given 
by the modular matrix $S^\sg$ as follows, 
\begin{equation}
   \langle \, \Phi^\la_{nm} (z,\bar z) \, \rangle_\La \ = \ 
   \frac{S^\sg_{\La \, \la}}{\sqrt{S^\sg_{0\, \la}}} \ 
   \frac{\delta_{nm}}{|z-\bz|^{2h_\la}} \ \ . 
\label{eq:CSCg} 
\end{equation}  
For the solution of the model, it would have been sufficient 
to present the couplings for the primary fields only, but we 
have decided to include all the fields from the list 
(\ref{eq:list1}). On the right hand side this gives rise to 
the trivial factor $\delta_{nm}$. Following a procedure 
suggested in \cite{DiVec,FFFS1} (see also \cite{MaMoSe1}), the 
localization region of the branes can be read off from the 
1-point functions (\ref{eq:CSCg}). The results of such an 
analysis confirm the findings of \cite{AlSc1} that Cardy 
type branes are localized along conjugacy classes 
\begin{equation} \label{ggeom}  
 C_L \ = \ \{ \, g \in G \ | \ g = u g_L u^{-1} \mbox{ for } 
   u \in G \ \} \ \ .
\end{equation}  
Here $g_L$ is some fixed group element which depends on the 
brane label $L$ (see e.g.\ \cite{FFFS1} for details).  
\medskip
 
Information on the spectrum of open string states on the 
branes (\ref{eq:CSCg}) is encoded in the fusion rules. More 
precisely, the space of open strings stretching between 
two branes $\La_1$ and $\La_2$ is contained in 
\begin{equation} \H_{\La_1}^{\sg; \La_2} \ = \ \bigoplus_\la \ 
 N_{\La_1\, \la}^{\sg; \La_2} \ \H^\sg_\la \ \  
\label{eq:OSdec} \end{equation}
where $N^\sg$ are the fusion rules of $\hg_k$. There is a 
boundary field associated with each state in this state 
space and one can show that the operator product expansion 
of any two such fields is determined by the fusing matrix 
\cite{Run1} (see also \cite{AlReSc1,FFFS1,FFFS2}).  
\smallskip

If we let $k$ tend to infinity while keeping the brane labels 
$L_1, L_2$ fixed, the space of ground states stays finite and
it is easy to identify it with $\Hom(V^{L_1},V^{L_2})$, i.e.\ 
with the space of linear maps between the two finite dimensional 
representation spaces $V^{L_1}$ and $V^{L_2}$ of the Lie algebra
$\sg$. For $L= L_1 = L_2$, the space $\Hom(V^L,V^L)$ comes 
equipped with a natural product (``matrix multiplication'') 
and it is exactly this product that one obtains from the OPE 
of open string vertex operators in the limit $k \rightarrow 
\infty$ \cite{AlReSc1}. Its non-commutativity can be nicely 
explained by the presence of a non-vanishing B-field on the 
branes.   
\smallskip     

After these remarks on the (non-commutative) geometry of 
branes in group manifolds we are prepared to review their 
low energy effective gauge theory. Consider
some configuration $P = \sum P_L (L)$ of Cardy-type branes 
on the group manifold which contains $P_L$ branes of type 
$(L)$ on top of each other. In the following we will not 
distinguish in notation between such a brane configuration 
$P$ and the associated representation $P$ of $\sg$. In 
particular, we shall denote by $V^P$ the reducible  
representation space $V^P = \sum P_L V^L$. It was shown in 
\cite{AlReSc2} that the effective action for the brane
configuration $P$ is given by a linear combination of a 
Yang-Mills and a Chern-Simons term for a set of fields 
$\tA_\a \in \End(V^P)$, 
\begin{equation}  \label{eq:effact}
  \cS_{P}\ =\  \cS_{{\rm YM}} + \cS_{{\rm CS}}\ = \
      \frac{1}{4}\ \tr \left( \tF_{\a \b} \ \tF^{\a \b} \right) 
     - \frac{i}{2}\  \tr \left( \tf^{\a \b \c}\; \tCS_{\a \b \c} \right)
\end{equation}
where we defined the `curvature form' $\tF_{\a\b}$ and some 
non-commutative analogue $\tCS_{\a\b\c}$ of the Chern-Simons form
by the expressions     
\begin{eqnarray} \label{eq:fieldstr}
  \tF_{\a \b}(\tA) & = &    
   i\, \tL_\a \tA_\b - i\, \tL_\b \tA_\a + i \,[ \tA_\a \astk \tA_\b] 
   +  \tf_{\a \b \c } \tA^\c \\[2mm]
   \label{eq:CSform}
  \tCS_{\a \b \c}(\tA) & = & \tL_\a \tA_\b \, \tA_\c 
                   + \frac{1}{3}\; \tA_\a \, [ \tA_\b \astk \tA_\c]
                   - \frac{i}{2}\; \tf_{\a \b \d}\; \tA^\d \, \tA_\c \ \ .
\end{eqnarray}
We have introduced the symbol $\tL_\a$ to denote the `infinitesimal 
translation' $\tL_\a \tA = [ P(t_\a), \tA]$ where $t_\a$ denote the 
generators of the Lie algebra $\sg$. Gauge invariance of 
(\ref{eq:effact}) under the gauge transformations 
$$ \tA_\a  \ \rightarrow \ \tL_\a \Lambda \ +\  i\, [\, \tA_\a\, ,\, 
    \Lambda\, ]  \ \ \ \mbox{ for } \ \ \ \Lambda \in \End (V^P)  
$$
follows by straightforward computation. Similar gauge theories 
on matrix (``fuzzy'') geometries \cite{Hop,Mad} have been studied 
before they were shown to appear in string theory (see e.g.\ 
\cite{GrKlPr3,GrKlPr2,GrKlPr1,WatWat,Kli,Madbook}).   
\medskip

{}From eq.\ (\ref{eq:effact}) we obtain the following equations of 
motion for the elements $ Q_\a := P(t_\a) + \tA_\a \in \End(V^P)$,   
\begin{equation} 
\label{eq:eom} 
\Bigl[\, Q^\a  \;\, , [\, Q_\a \, , \,  Q_\b\, ] \, - \,  
 i\, \tf_{\a \b \c}\, Q^\c \,\Bigr]  \ = \ 0 \ \ . 
\end{equation} 
Solutions of these equations (\ref{eq:eom}) describe possible 
condensates of our brane configuration $P$. There exists one type
of solutions that is particularly interesting. Obviously, we can 
satisfy the equations (\ref{eq:eom}) by choosing $Q_\a$ to be any 
$\dim(V^P)$-dimensional representation of the Lie algebra $\sg$.
The associated solution is then given by $\tA_\a = Q_\a - P(t_\a)$. 
As it was argued in \cite{AlReSc2}, this solution describes a  
process of the form 
$$ ( P )  \ \stackrel{\tA = Q - P }{\longrightarrow} \ 
   ( Q ) \ \ . $$
Support for this statement comes from both the open string 
sector and the coupling to closed strings (see \cite{AlReSc2}). 
On the one hand, we can compare the tension of D-branes in the 
final configuration $Q$ with the value of the action $\cS_P(\tA)$  
at the classical solution $\tA$. On the other hand, we can look 
at fluctuations around the chosen solution and compare their 
dynamics with the low energy effective theory $\cS_{Q}$ of the 
brane configuration $Q$. In formulas, this means that 
\begin{equation}\label{eq:fluctuation}
\cS_{P} (\tA+\delta \tA)\ \overset{!}{=}\  \cS_{P} (\tA) + 
\cS_{Q} (\delta \tA) \ \ \ \mbox{ with } \ \ \
\cS_P(\tA) \ \overset{!}{=} \ \ln \frac{g_{Q}}{g_P} \ \ . 
\end{equation}
The second requirement expresses the comparison of tensions in 
terms of the g-factors \cite{AffLud1} of the involved  conformal 
field theories (see e.g.\ \cite{AlReSc2} for more details). All 
equalities must hold to the order in $(1/k)$ that we used when 
we constructed the effective actions.  
\smallskip

These results imply that condensation processes of Cardy type 
brane configurations $P$ on group manifolds possess only one 
invariant: the dimension $\dim (V^P)$ of the representation 
$P$. We can easily identify this invariant with the D0 brane
charge. In fact, a particular initial configuration is given 
by the representation $P = P_0 [0]$, i.e.\ by choosing the 
trivial representation of $\sg$ with multiplicity $P_0$. It 
corresponds to a configuration in which $P_0$ point-like branes
are placed on top of each other at the group unit $e \in G$. 
Now we are advised to pick any $\dim (V^P) = P_0$-dimensional 
representation $Q$ of $\sg$. The latter decomposes into a 
sum of irreducible representations $Q = \sum Q_L [L]$. 
If $Q$ is irreducible, the final state contains a single 
extended brane of charge $P_0$. A well known example of this
phenomenon is the formation of spherical branes on $S^3 
\cong$ SU(2) which was discussed extensively in the past 
(see \cite{AlReSc2} and also  \cite{HasKra},\cite{HiNoSu}). 
Similar effects have been described for branes in RR-background 
fields \cite{Mye}. The advantage of our scenario with 
NSNS-background fields is that it can be treated in 
perturbative string theory so that string effects may be 
taken into account (see \cite{FrSc1}).

\section{Boundary coset models} \label{sec:bcm}
\setcounter{equation}{0} 
From now on let $H \subset G$ denote some simple simply 
connected subgroup of $G$. We want to study the associated 
$G/H$ coset model. A more precise formulation of this theory 
requires a bit of preparation (more details can be found e.g.\ 
in \cite{DiFrabook}). We shall label the sectors $\H^\sh_{\la'}$ 
of the affine Lie algebra with $\hh_{k'}$ labels $\la' \in 
\cJ^\sh_{k'}$. Note that the sectors of the numerator theory 
carry an action of the denominator algebra $\hh_{k'} \subset 
\hg_k$ and under this action each sector $\cH^\sg_\la$ 
decomposes according to 
$$ \H^\sg_\la \ = \ \bigoplus_{\la'} \ \H_{(\la,
     \la')} \otimes \H^\sh_{\la'}\ \ . $$
Here we have introduced the infinite dimensional spaces 
$\H_{(\la,\la')}$ which we want to interprete as 
sectors of the coset chiral algebra. The latter is usually 
hard to describe explicitly, but at least it is known to 
contain a Virasoro field with modes
\begin{equation}\label{eq:cosetvirasoro}
 L_n \ = \ L^\sg_n - L^\sh_n \ \ . 
\end{equation}
One may easily check that they obey the usual exchange 
relations of the Virasoro algebra with central charge given 
by $c = c^\sg -c^\sh$. 
\smallskip

Note that some of the spaces $\H_{(\la,\la')}$ may vanish  
simply because a given sector $\H^\sh_{\la'}$ of the 
denominator theory may not appear as a subsector in a 
given $\H^\sg_\la$. This motivates to introduce the set  
$$ \cE \ = \ \{ \, (\la, \la') \in \cJ^\sg_k \times 
   \cJ^\sh_{k'}\, | \, \cH_{(\la, \la')} \neq 0\, 
    \} \ \ . $$ 
Elements of $\cE$ do not yet label sectors of the coset 
models. In fact, different elements of $\cE$ may correspond 
to the same sector, i.e.\ there is an equivalence relation 
$$ (\la,\la') \sim (m,m') \ \Leftrightarrow 
 \ \H_{(\la,\la')} \cong \H_{(m,m')} \ \ . $$
At this point we want to make one assumption, 
namely that all the equivalence classes we find in $\cE$ 
contain the same number $N_{0}$ of elements. This holds true for many 
important examples and it guarantees that the sectors of the 
coset theory are simply labeled by the equivalence classes, 
i.e.\ $\cJ = \cE/ \sim$. \footnote{For more general cases, there are 
further sectors that cannot be constructed within the sectors 
of the numerator theory.} It is then also easy to spell out 
explicit formulas for the fusion rules and the S-matrix 
of the coset model. These are given by
\begin{eqnarray} 
N^{(k,k')}_{(\la,\la')\, (m,m')} & = & 
 \sum_{(n ,n ')\sim (k , k ')}   
N^{\sg; n}_{\la\, m}\, N^{\sh;n'}_{\la'\,m'} \ \ , 
\label{eq:cosN} \\[2mm] 
S_{(\la,\la')(m,m')} & = & 
 N_{0}\, S^\sg_{\la\, m} \ \bar S^{\sh}_{\la'\, m'} \ \ 
\label{eq:cosS} \end{eqnarray} 
where the bar over the second S-matrix denotes complex 
conjugation. 
\smallskip

Let us note that $(\la,\la')$ is an element of $\cE$ if (but not
only if) the representation $\la'$ of the finite dimensional
Lie algebra $\sh$ appears as a sub-representation of the 
representation $\la$ for $\sg$. The equivalence classes 
of such special pairs form a subset $\cJ^r \subset \cJ$.  
Sectors in the subset $\cJ^r$ are also distinguished 
because the conformal dimension of their ground state 
satisfies the equality  
$$ h_{(\la,\la')} \ = \ h^\sg_\la - h^\sh_{\la'} + n 
$$
with $n = 0$. For other pairs $(\la,\la') \in \cJ$, $n$ 
is a (non-vanishing) positive integer. This means in 
particular, that fields associated with the sectors in 
$\cJ \setminus \cJ^r$ necessarily have conformal dimension 
$h,\bar h > 1$ and hence they are irrelevant. We shall see 
in Appendix A that the sectors labeled by elements of 
$\cJ^r$ play a special role when we analyse the brane 
geometry.  
\medskip

The boundary theories we are going to look at are 
associated with the diagonal modular invariant bulk 
partition function, 
$$ Z(q,\bar q) \ = \ \sum_{(\la,\la')} \ \chi_{(\la,\la')}(q) 
   \chi_{(\la,\la')}(\bar q) \ \ . $$
We want to impose trivial gluing conditions along the 
boundary in which each left moving chiral field of the 
coset theory is glued to its right moving partner. Under 
this condition, the associated boundary theories can be 
constructed using Cardy's solution \cite{Car}. It asserts 
that the model has as many boundary conditions as there are 
sectors of the coset algebra. We will label them with  
$(\La,\La') \in \cJ$. For a given theory $(\La,
\La')$ the couplings of closed string modes to the boundary 
are essentially given by the S-matrix, i.e.\ 
\begin{equation} \label{cos1pf}  
 \langle \, \phi_{(\la,\la')} (z,\bz) \, \rangle_\La \ = \ 
 \frac{S^\sg_{\la\,\La} \bar S^{\sh}_{\la'\,\La'}}{\sqrt{ 
 S^\sg_{0\, \la} \bar S^{\sh}_{0\la'}}} \ 
\frac{1}{|z-\bar z|^{2h_{(\la,\la')}}} \ \ . 
\end{equation} 
Here we have used the explicit formula for the S-matrix of
the coset theory that was given in eq.\ (\ref{eq:cosS}). Let 
us also write down the spectrum of open string modes 
stretching in between two branes $(\La_1,\La'_1)$ and
$(\La_2,\La'_2)$, 
\begin{equation} Z_{(\La_1,\La'_1)}^{(\La_2,\La'_2)} (q) \ = \ 
 \sum_{(l,l')} \ N^{\sg; \La_2}_{\La_1\, l} \ N^{\sh; 
  \La'_2}_{\La'_1\, l'}   \ \chi_{(l,l')} (q) \ \ . 
\label{eq:cosOSdec} \end{equation} 
This formula involves the fusion rules of the coset model 
that were spelled out in eq.\ (\ref{eq:cosN}). Let us point 
out that we can think of these elementary Cardy branes as 
being labeled by a set of irreducible representations 
$[L,\bL']$ of $\sg \oplus \sh$. Here $\bL'$ denotes the 
representation conjugate to $L'$. This way of associating 
a representation $[L,\bL']$ to the brane configuration 
$(L,L')$ turns out to be rather convenient. More complicated 
brane configurations $P$ involve reducible representations of 
the same Lie algebra. As in the case of branes on group 
manifolds we will often identify brane configurations $P$ 
with the associated representation of $\sg \oplus \sh$.     
\smallskip

The geometry of the Cardy type branes in coset models was 
recently uncovered in \cite{Gaw2} (see also \cite{EliSar}). 
To describe the answer we need some more notation. Recall 
first that geometrically the quotient $G/H$ is formed with 
respect to the adjoint action of $H$ on $G$, i.e.\ two 
points on $G$ are identified if they are related by 
conjugation with an element of $H \subset G$. We denote 
the projection from $G$ to the space $G/H$ of $H$ orbits
by $\pi^{G}_{G/H}$. Furthermore, we use $C^G_L$ to refer 
to the conjugacy class of $G$ along which the brane with 
$L$ is localized and similarly for $C^H_{L'}$. The latter 
is a conjugacy class in $H$. Through the embedding of $H$ 
into $G$, we can regard it as a subset of $G$. Now we 
construct the set $C_{(L,L')}$ of all elements in $G$ 
which are of the form $ u v^{-1}$ where $u \in C^G_L$ 
and $v \in C^H_{L'}$. This set is left invariant by 
conjugation with elements of $H$ and hence it can be 
projected down to $G/H$. The claim of \cite{Gaw2} is that 
the brane $(L,L')$ is localized along the resulting subset 
$C^{G/H}_{(L,L')}$ of $G/H$, 
\begin{equation} \label{cosgeom} 
 C^{G/H}_{(L,L')} \ = \ \pi^{G}_{G/H}\left(\, 
    C^G_L \ (C^H_{L'})^{-1} \, 
    \right) \ \subset \ G/H \ \ . 
\end{equation} 
We shall extract this result from an analysis of the 
1-point function in Appendix A.

\section{Coset branes and fuzzy gauge theories}
\setcounter{equation}{0} 
In this section we will discuss the  effective non-commutative 
gauge theory that describes the dynamics of branes in coset 
theories. Given some configuration of coset branes, i.e.\ a 
representation of $\sg \oplus \sh$,  we construct some parent 
action which is essentially identical to the field theory 
(\ref{eq:effact}).  The effective field theory of coset branes 
is then obtained by a suitable reduction. Our construction 
can be derived from conformal field theory. While we explain 
most of this in Appendix B, we provide the derivation for
special brane configurations of the form $P = \sum P_L (L,0)$  
in the second subsection below.

\subsection{Construction of the effective field theory}
\def\bL{{\bar L}}

We have reviewed the effective action for branes on group manifolds
in Section \ref{sec:review}. The result has been discussed for a 
WZW model involving a single affine Lie algebra $\hg_k$. For our 
purposes below, we need the action for cases where the underlying 
affine Lie algebra is a direct sum of algebras with different 
levels $k_{r}$. From the resulting action we will then obtain 
the effective field theory of coset branes by reduction. 
\smallskip 

A coset model involves two chiral algebras $\hg$ and $\hh$ 
in the numerator and the denominator, respectively. In general,
these possess decompositions of the form $\hg = \hg_{1}\oplus 
\cdots \oplus \hg_{R}$ and $\hh = \hh_{1} \oplus \dots \oplus 
\hh_{R'}$ with possibly different levels $k_1, \dots, k_R; 
k'_1 , \dots , k'_{R'}$ appearing in each summand. We will 
study a regime of the model in which some of the levels are 
sent to infinity while others stay finite. Let us assume that 
the decompositions above have been arranged such that $k_1, 
\dots, k_S$ and $k'_1, \dots k'_{S'}$ become large. 

In this limiting regime we intend to study Cardy type brane 
configurations $P = \sum P_{L\, L'} (L,L')$ where $L,L'$ 
are multi-labels of the form $L = (L_1, \dots, L_S, 0, \dots,
0)$ and $L' = (L'_1, \dots, L'_{S'},0, \dots, 0)$ in which 
the representation labels for the small directions are chosen 
to be trivial\footnote{In the limit of large $k$ the theory 
is essentially independent of the labels $L_{S+1}, \dots, L_R$, 
$L_{S'+1}, \dots, L_{R'}$}. As we explained before, such a brane 
configuration gives rise to a representation $P = \sum P_{L\, L'} 
[L, \bL']$ of the Lie algebra $\sg \oplus \sh$.          
\smallskip 

The field theory we are going to spell out now involves a number
of gauge fields $\tA_\a$ where $\a$ label a basis in $\sg \oplus 
\sh$, i.e.\ it runs through the values $1, \dots, \dim \sg + \dim 
\sh$. The gauge fields 
$\tA_\a$ are elements of the space $\End(V^P)$ which depends 
on the choice of our initial brane configuration $P$. Let us also
introduce the derivatives $\tL_\a$ as follows 
\begin{equation} 
\tL_\a \tA \ = \ \left\{ \begin{array}{ll} \ [\, P(t_\a)\, , \, \tA\, ] 
  \ \ & \mbox{ for } \ \ \ \a  \leq \dim \sg \\[2mm] 
   i\, [\, P(t_\a)\, , \, \tA\, ] 
  \ \ & \mbox{ for } \ \ \ \a > \dim\sg  \end{array} \right. 
\ \ . \end{equation}          
Note that we have absorbed an extra factor $\sqrt{-1}$ into the 
definition of $\tL_\a, \a > \dim \sg$. This will turn out to be rather 
convenient in the following. In these notations, we are now able
to introduce the following action, 
\begin{equation}
\cS^{\rm WZW}_{P} (\tA ) \ =\ \cS_{\rm YM} (\tA )+\cS_{\rm
CS} (\tA)\ = \ \frac{1}{4}\tr (\tF_{\alpha \beta }\tF^{\alpha \beta
})-\frac{i}{2}\tr (\tilde \tf ^{\alpha \beta \gamma }\tCS_{\alpha \beta
\gamma }) \;
\label{pWZWact} \end{equation}
where $\tilde \tf^{\a\b\c} \ = \ \tf^{\a \b \c}$ if $\a,\b, \c \leq 
\dim \sg$ and  $\tilde \tf^{\a\b\c} \ = \ i  \tf^{\a \b \c}$
otherwise. 
Let us also note that the indices $\a$ are raised and lowered using the
open string metric
\begin{equation} \label{eq:openstringmetric}
 G^{\alpha\, \alpha'} \ =\ \frac{2}{k(\alpha)} \ \delta^{\a\, \a'}
 \;. 
\end{equation} 
Here, the function $k(\a)$ has been introduced such that it takes 
the value $k_r$ (or $k'_{r'}$) if $\a$ refers to a basis element 
in the Lie-algebra $\sg_r$ (or $\sh_{r'}$). 
\medskip 

We use effective action $S^{WZW}$ as a master theory from which we 
descend to the effective description of branes in coset conformal 
field theory. For the reduction it is convenient to switch to a 
new basis of $\sg \oplus \sh$ which makes reference to the embedding 
$\sh \subset \sg$. We shall employ $a = 1, \dots, \dim\sg - \dim \sh$ 
when we label directions perpendicular to $\sh \subset \sg$ while 
labels $i = 1,\dots, \dim \sh$ and $\im = 1 , \dots, \dim \sh$ 
stand for directions along $\sh \subset \sg$ and $\sh$, respectively. 
\label{sec:effaction}

The idea is now to perform a reduction of the theory (\ref{pWZWact}) 
by imposing the following constraints \vspace*{-4mm} 
\begin{equation}
\begin{array}{c}
 \tA_i\ =\ \tA_{\im}\ =\ 0  \\[2mm]
 (i\tL_i+\tL_{\im})\tA_a+ \tf_{ia}^b\tA_b \  = \   0\ .
\end{array}
\label{eq:genconstraint}
\end{equation} 
These conditions allow to rewrite the effective action in the form 
\begin{equation}\label{eq:cosetaction}
\cS_{P} (\tA )\ =\ \frac{1}{4}\tr (\tF_{ab}\tF ^{ab})
-\frac{i}{2}\tr (\tf^{\alpha\beta\gamma}\tCS_{\alpha\beta\gamma})
\end{equation}
which, together with the constraints \eqref{eq:genconstraint}, 
determine the brane dynamics in coset models. The field strength 
$\tF $ and the Chern-Simons form $\tCS $ are defined as before in 
\eqref{eq:fieldstr},\eqref{eq:CSform}, but with $\tA _{i},\tA_{\im}$ 
set to zero. Formulas (\ref{eq:cosetaction},\ref{eq:genconstraint}) 
constitute the central result of this section. 

\subsection{Derivation of the action for $P = \sum P_L (L,0)$}

The effective theory we proposed in the previous subsection can 
be derived from boundary conformal field theory. We want to
explain this here at least for a restricted set of brane 
configurations $P = \sum P_L (L,0)$ in which the denominator 
labels are all set to zero, i.e.\ $L'_r= 0$. This implies that 
the state space of the configuration contains only sectors of 
the form $\H_{(l,0)}$ and it simplifies the discussion 
considerably. For the general case the reader is referred to 
Appendix B where we sketch the main ideas of the derivation. 
\smallskip 
 
Since the contribution of the denominator theory is trivial, 
we can restrict our attention to the numerator theory with 
chiral algebra $\hg$. This theory has a state space of the 
form 
$$ \H^P \ = \ \bigoplus \ P_L\  N^{\sg; L}_{L l} \ \H^\sg_l   $$ 
where $\H_l$ denotes the state space of the sector $l$ of 
the chiral algebra $\hg$. To get rid of excitations in the 
direction of $\sh \subset \sg$, we have to impose the 
conditions 
\begin{equation}\label{eq:restriction}
J^{i}_{n}\ \psi \ =\ 0 \quad \mbox{ for } \ \  n\geq 0 \ \ 
\mbox{ and } \ \  i = 1, \dots, \dim \sh 
\end{equation}
on $\psi \in \cH^{\sg}_{l}$.  The subspace of states that 
solves these constraints is given by 
\begin{equation}
 \bigoplus \ P_L \ N^{\sg; L}_{L l}\  \cH_{(l,0)}\, \otimes \,  
    |0\rangle^\sh  \ \simeq \  \bigoplus P_L \ N^{\sg; L}_{L l} 
  \  \cH_{(l,0)} \ \subset \ \H^P \ \ . 
\end{equation}
Omitting the vector $|0\rangle^{\sh}$ is justified here because 
it has vanishing conformal dimension and the operator product 
expansions of the associated identity field are all trivial.
Hence, the isomorphism indicated by $\simeq$ is canonical, i.e.\ 
it preserves all the structure that we need to compute the 
effective theory. Obviously, our assumption $L'_r=0$ is crucial 
at this point.  
\smallskip

These observations give a a good handle to compute the effective 
action for the coset model from the known effective action for 
the $G$ WZW-model. All we have to do is to implement the restrictions 
\eqref{eq:restriction} described above directly on the fields 
$\tA_\a$ of the effective theory. This gives  
\begin{eqnarray}\label{eq:constraint1}
  \tA^{i} \ = \  2i\alpha' \tf^{i\gamma \a }\tL_{\gamma }
\tA_{\alpha } \quad
& & \mbox{ for all } \  \; i \\[2mm] 
\label{eq:constraint2}
i \tL_{i}\tA_{\beta }
+ {\tf_{i\b }}^{\alpha }\tA_{\alpha } \ = \  0 \quad
\ \  & & \mbox{ for all }\ \  \; i,\; \beta \; 
\end{eqnarray} 
where the first constraint follows from eq.\ (\ref{eq:restriction}) 
for $n> 0$ and the second constraint is obtained with $n=0$. By using 
\eqref{eq:constraint1} and \eqref{eq:constraint2} for $\beta =j$, 
we can express $\tA_{i}$ through $\tA_{a}$. Thus we eliminate $\tA_{i}$
from the action and are left with an action only containing $\tA_{a}$
subjected to the condition 
\begin{equation}\label{eq:constraint2'}
i \tL_{i}\tA_{b }+ {\tf_{i b }}^{a }\tA_{a } \ = \ 0\quad
\mbox{ for all } \ \ \  i,b \; .
\end{equation}
Furthermore it turns out that the terms in the action coming from the
$\tA _{i}$ are strongly suppressed against other terms. This is
suggested already by the appearance of $\alpha'$ in equation  
\eqref{eq:constraint1} and it allows us to neglect these
terms in the action to leading order so that $A_i =0$. The 
resulting theory agrees with the prescription given in 
Subsection 4.1. 

\section{Solutions and Condensation Processes} 
\setcounter{equation}{0} 
Having found the effective theory (\ref{eq:cosetaction},
\ref{eq:genconstraint}) of coset branes we shall now proceed 
to discuss a large class of solutions and their interpretation 
as condensation processes. In \cite{FrSc2} we reported on the 
results from the reduction procedure in the parafermion case, 
${\rm SU} (2)/{\rm U} (1)$. This section will be a generalization 
to arbitrary fixed point free coset models.

\subsection{Solutions}\label{sec:solutions}

To obtain the equations of motion we vary the action
\eqref{eq:cosetaction} under the constraint \eqref{eq:genconstraint}. 
It is easy to see that the variation vanishes away from the 
configurations fulfilling the constraints so the resulting 
equations are the same as in the unconstrained problem,
\begin{equation}\label{eq:coseteom}
\tL^{\alpha}\tF_{\alpha b}+[\tA^a,\tF_{ab}]\ =\ 0 \ \ . 
\end{equation}
Together with the constraints \eqref{eq:genconstraint}, eqs.\ 
(\ref{eq:coseteom}) determine the dynamics.

Consider a configuration $P=\sum_{L,L'}P_{L\,L'} (L,L')$. The fields
$\tA_{a}$ are matrices on which we can act with derivatives
$\tL_{\alpha }$. These can be expressed through commutators with the 
matrices $P_\alpha  = P(t_\a)$ given by the corresponding
representation $P$. Suppose now that we found a
decomposition $P_{\alpha}= Q_{\alpha }- Q'_{\alpha }$ \footnote{where 
$Q'$ and $Q$ are zero in 'small' directions, i.e.\ in directions 
corresponding to a small level} such that $Q$ is a representation 
of $\LA{g}\oplus\LA{h}$, 
\begin{eqnarray*}
[\, Q_{\alpha }\, ,\, Q_{\beta }\, ] & =&  i\ {\tf_{\alpha \beta}}^{\gamma}
\, Q_{\gamma }\ , \\[2mm]
\makebox[0cm]{\hspace*{-3cm}and} Q'_{i}\, +\, Q'_{\im} & = & 0 \ \ .
\end{eqnarray*}
Then 
\begin{equation}\label{eq:solution}
\tA_{a}\ =\ Q_a - P_a \ = \  Q'_{a}
\end{equation}
is a solution fulfilling the equations of motion \eqref{eq:coseteom} and 
the constraints \eqref{eq:genconstraint}. This can be verified in a 
straightforward computation.
Note that this construction generalizes the one that we sketched for
the case of branes on group manifolds. As we will show in the next
subsection also the interpretation of the solution is analogous: The 
solution describes a brane configuration given by the representation $Q$. 
The main difference is that 
for a non-trivial denominator, we are not free to choose any 
representation $Q$ of $\sg\oplus\LA{h}$ but 
have to satisfy the extra condition 
that the solution vanishes in the diagonal 
combination $Q'_i+Q'_{\im}$.
The latter becomes trivial for group manifolds
since the set of directions $i$ along the denominator is empty in 
this case.    

Let us comment on the meaning of this extra condition which is equivalent to
\begin{equation*}
P_i\, +\, P_{\im}\, = \, Q_i \, + \, Q_{\im} \ ,
\end{equation*}
saying that $P$ and $Q$ are isomorphic as representations of the
diagonally embedded $\LA{h}_{\rm diag}\subset \LA{g}\oplus\LA{h}$. 
Using the identification of the solution with a condensation   
process we see that
\begin{enumerate}
\item all processes we found leave the diagonal part $P|_{\LA{h}_{\rm diag}}$
invariant
\item any two brane configurations $P,Q$ satisfying 
$P|_{\LA{h}_{\rm diag}}\sim Q|_{\LA{h}_{\rm diag}}$ are connected 
through a condensation process.
\end{enumerate}

It is finally worth noticing that there is a distinguished solution 
which exists for any configuration $P$, namely we can choose 
\begin{equation*}
Q'_a\ =\ -P_a\ , \ Q'_i =\ -P_i\ ,\ Q'_{\im}\ =\ P_i \ \   
\end{equation*}
The solution relates $P$ to the following configuration $Q$ 
\begin{equation*}
Q_a\ =\ 0\ ,\ Q_i\ =\ 0\ ,\  Q_{\im}\ =\ P_i +P_{\im}\ .
\end{equation*}
Since $Q$ describes a superposition of $(0,L')$ branes, we have 
just shown that any brane configuration $P$ can be related by a 
condensation process to a superposition of $(0,L')$-branes. 
This is in complete agreement with an investigation of closed string
couplings in \cite{LerWal} for a family of Kazama-Suzuki coset models. There it
was observed that the $(0,L')$ branes provide a basis for the lattice 
of Ramond-Ramond charges.

\subsection{Interpretation of the solutions}\label{sec:interpretation}
The described solutions can be identified as a condensation process 
that leads either to or away from the initial brane configuration
$P$. Let us reformulate the proposal for the configuration that is 
associated with the solution $Q'$. The set of matrices $Q_{\alpha }$ 
form a representation of $\LA{g} \oplus \LA{h}$. This
representation can be decomposed into irreducible sub-representations
$\bigoplus Q_{L\, L'} V_{L}\otimes V_{\bar{L}'}$. We now claim that this
decomposition describes the brane configuration $\sum Q_{L\,
L'}(L,L')$ we are looking for. Whether the process is a flow to or 
from this configuration depends on $\cS (Q')$ being negative or
positive, respectively. As evidence for this interpretation we will 
give here an analysis of D-brane tensions and of fluctuation spectra. 
What we will show can be summarized in the formula
\begin{equation}
\cS_{P} (Q'+ \tA)\ =\  \cS_{P} (Q') + 
\cS_{Q} (\tA) \ \ \ \mbox{ with } \ \ \
\cS_P(Q') \ = \ \ln \frac{g_{Q}}{g_P} \ \  
\end{equation}
analogous to \eqref{eq:fluctuation}.

For the calculations it is useful to relate our reduced action to the 
unreduced WZW action \eqref{pWZWact}. It can be shown that
\begin{equation}
\cS_P(\tA_a) = \cS_P^{\rm WZW}(\tA_a,\tA_i=Q'_i,\tA_{\im}=iQ'_{\im})
\end{equation}
for any $Q'_i=-Q'_{\im}$ belonging to a solution \eqref{eq:solution} and
for all $\tA_a$ fulfilling the constraint \eqref{eq:genconstraint}. Note 
the appearance of a factor of $i$ because of our conventions for 
the $\LA{h}$-part.

Now let us expand our coset action for a solution $Q'_a$ using the 
result for WZW models \eqref{eq:fluctuation}.
\begin{eqnarray*}
\cS_P(Q'_a+\tA_a) &= &\cS_P^{\rm WZW}(Q'_a+\tA_a\, ,\, Q'_i\, ,\, iQ'_{\im})\\[2mm]
& = & \cS_P^{\rm WZW}(Q'_a+\tA_a\, ,\, Q'_i+\tA_i\, ,\, iQ'_{\im}+\tA_{\im})|_{\tA_i=\tA_{\im}=0}\\[2mm]
& = & \cS_P^{\rm WZW}(Q'_a\, ,\, Q'_i\, ,\, iQ'_{\im})\ +\ 
\cS_Q^{\rm WZW}(\tA_a\, ,\, \tA_i\, ,\, \tA_{\im})|_{\tA_i=\tA_{\im}=0}\\[2mm]
&=& \cS_P^{\rm WZW}(Q'_{\alpha}) \ +\  \cS_Q(\tA_a) \ \ .
\end{eqnarray*}
This confirms our result that the fluctuations around the solution $Q'_a$
are governed by the action corresponding to the brane configuration $Q$.
To complete this argument we note that the 
constraint \eqref{eq:genconstraint} for the $\tA_a$ in the 
$P$-configuration is the same as in the $Q$ configuration as
\begin{equation*}
i\tL_i+\tL_{\im}\ =\ i[P_i+P_{\im},\ \cdot\ ]\ =\ i[Q_i+Q_{\im},\ \cdot\ ]\ .
\end{equation*}

In the remaining part of this section we will show that the D-brane tensions
are reproduced correctly by our solution, i.e.\ 
\begin{equation}\label{eq:tension}
{\rm ln} \frac{\sum Q_{L\,L'}g_{L,L'}}{\sum P_{L\,L'}g_{L,L'}}\  = \ 
\cS (Q'_a)\ .
\end{equation}
in some order of $1/k$.
The g-factors are defined by
\begin{equation}
g_{L,L'}\ =\ \frac{S_{(L,L') (0,0)}}{\sqrt{S_{(0,0) (0,0)}}}
\end{equation}
with the help of the coset S-matrices.
The coset S-matrices are up to some constant factor just products of
S-matrices of the involved affine Lie algebras in the numerator and
the denominator, therefore
\begin{equation}\label{eq:gfactors}
\frac{\sum Q_{L\,L'}g_{L,L'}}{\sum P_{L\,L'}g_{L,L'}}\ =\ \frac{\sum
Q_{L\,L'}\prod S^{r}_{L_{r}0}\prod S^{r'}_{L'_{r'}0}}{\sum
P_{L\,L'}\prod S^{r}_{L_{r}0}\prod S^{r'}_{L'_{r'}0}}\ \ .
\end{equation}
As we are performing a perturbative analysis in $1/k$ we need
asymptotic expressions for S-matrices. Using expressions from
\cite{DiFrabook} we find
\begin{equation}
S_{\la 0}\ =\ N (k) \ \dim (\la ) 
\Bigl(1-\frac{\pi^{2}}{6 (k+g^{\vee })^{2}}g^{\vee }
C_{\la }+\mathcal{O} \bigl(\frac{1}{k^{4}}\bigr)\Bigr)
\end{equation}
where $N (k)$ is some factor independent of $\la $, $C_{\la }=
(\la ,\la +2\rho )$ is the quadratic Casimir, and $g^{\vee }$
is the dual Coxeter number.
We insert this expression into \eqref{eq:gfactors} and obtain
\begin{eqnarray*}
\lefteqn{\frac{\sum Q_{L\,L'}g_{L,L'}}{\sum P_{L\,L'}g_{L,L'}} \ =\ 
\frac{\sum Q_{L\,L'}\dim_{L,L'}}{\sum P_{L\,L'}\dim_{L,L'}}}\\[2mm]
&&-\frac{\pi ^{2}}{6}
\frac{1}{\sum Q_{L\,L'}\dim_{L,L'}}\sum
Q_{L\,L'}\dim_{L,L'}\Bigl[\sum_{r}\frac{g_{r}^{\vee
}C^{r}_{L_{r}}}{(k_r+g_{r}^{\vee })^{2}}+
\sum_{r'}\frac{g_{r'}^{\vee
}C^{r'}_{L'_{r'}}}{(k_{r'}+g_{r'}^{\vee })^{2}}\Bigr] \\[2mm]
&&+
\frac{\pi ^{2}}{6}
\frac{1}{\sum P_{L\,L'}\dim_{L,L'}}\sum
P_{L\,L'}\dim_{L,L'}\Bigl[\sum_{r}\frac{g_{r}^{\vee
}C^{r}_{L_{r}}}{(k_{r}+g_{r}^{\vee })^{2}}+
\sum_{r'}\frac{g_{r'}^{\vee
}C^{r'}_{L'_{r'}}}{(k_{r'}+g_{r'}^{\vee })^{2}}\Bigr]
+ \mathcal{O}\bigl(\frac{1}{k^{4}}\bigr)\ .
\end{eqnarray*}
We now want to check the condition \eqref{eq:tension} for our proposed
interpretation. The value of the action is
\begin{eqnarray*}
\cS_P(Q'_a) & =&\cS_P^{\rm WZW}(Q'_{\alpha})\\
& = & \frac{1}{12} {\tf_{\alpha \beta }}^{\gamma }
\tf ^{\alpha \beta \delta  }\tr (P_{\gamma }P_{\delta
}-Q_{\gamma } Q_{\delta }) \ \ .
\end{eqnarray*}
Remembering that indices are raised and lowered with the help of the
$k$-dependent open string metric \eqref{eq:openstringmetric}, we can see
that this result is of order $1/k^{2}$. Since in our case we have 
\begin{equation*}
\sum P_{L\, L'}\dim [L,L']\ =\ \sum Q_{L\, L'}\dim [L,L'] \; ,
\end{equation*}
the left hand side of \eqref{eq:tension} is also of order
$1/k^{2}$. It is then straightforward to show that indeed
\eqref{eq:tension} is fulfilled to this order.

Let us briefly mention that it may happen that the action vanishes 
in the order $1/k^{2}$. This is the case if all 'large' directions
\footnote{by large directions we mean those which belong to a large 
level $k_{r}$} that are used in the construction of the solution are 
divided out. We will encounter such a case in the example of the 
minimal models. However, it can be shown that in this case the 
relation \eqref{eq:tension} is fulfilled also in the order 
$1/k^{3}$.

\section{Examples: Parafermions and minimal models}
\setcounter{equation}{0} 
In this final section we want to illustrate our very general 
results in three simple examples. It will become clear that the 
solutions we have constructed above are capable of describing 
brane processes with very different geometrical manifestations. 
In the case of minimal models we will recover the processes 
found in \cite{ReRoSc} among our solutions and we shall now 
provide a very nice geometrical picture for them.  

\subsection{Parafermions}
Let us start by reviewing the construction of parafermion
theories as $\widehat{\rm su}(2)_{k}/\widehat{\rm u}(1)_{k}$ cosets. 
The free bosonic U(1) theory is embedded such that its current gets 
identified with the component $J^{3}$ of the SU(2) current.
\smallskip

The numerator theory has sectors $\cH^{\rm su(2)}_{l}$ where $l = 
0,1, \dots,k$, the sectors $\cH^{\rm u}_{m}$ of the denominator algebra 
$\widehat{\rm u} (1)_{k}$ carry a label $m = -k+1, \dots , k$.
We can label
the sectors $\cH_{(l,m)}$ of the coset model by pairs $(l,m)$ of
numerator and denominator levels. The possible pairs $(l,m)$ are
restricted by a selection rule forcing the sum $l+m$ to be even. 
Furthermore some pairs label the same sector so that we have to
identify the pairs $(l,m)\sim (k-l,m+k)$. Here we take the label $m$
to be $2k$-periodic. Note that this field identification has no fixed
points. \smallskip

Now we want to apply our general formalism to formulate the effective
action for the parafermion branes. Let us illustrate here only the case where
the branes have trivial label in the denominator part.
We start with the effective action
for the $\widehat{\rm su} (2)$-WZW model involving three fields 
$\tA_{1},\tA_{2},\tA_{3}$. Our brane configuration $P=\sum P_{L\, 0}[L,0]$
determines the derivatives $\tL_{\alpha}=[P(t_{\alpha}),\ \cdot\ ]$.
The constraint \eqref{eq:genconstraint} reads in the parafermion case
\begin{equation}\label{eq:pfconstraint2}
i\tL_{3}\tA_{a}
+{\tf_{3a }}^{b }\tA_{b } \ = \  0 \quad
a,b=1,2 \quad ,
\end{equation}
$\tA _{3}$ is set to zero. Eventually we arrive at the effective
action for the coset theory. The result is 
\begin{equation}\label{eq:pfaction}
\cS (\tA_{1},\tA_{2}) \ = \  \frac{1}{4}\tr (\hat{\tF}_{ab}\hat{\tF}^{ab})
\end{equation}
where $a=1,2$ and $\hat{\tF}_{ab}=i\tL_{a}\tA_{b}-i\tL_{b}\tA_{a}+
i[\tA_{a},\tA_{b}]$. Obviously, there is no Chern-Simons like term 
in this case simply for dimensional reasons. 
\smallskip

Let us now analyze the effective theory on a single $(L,0)$ brane. 
For $L>0$ we find a solution of the form described in \ref{sec:solutions}
given by the following non-constant field
\begin{equation}\label{eq:pfsolution}
\tA_{a}\ =\ -P_{a} \ = \ - P(t_a) 
\end{equation}
which is rather easy to check here for the parafermions.

If we insert this solution into the action \eqref{eq:pfaction}
we find a positive value, indicating that the brane is the decay
product of some configuration with a higher mass.
This configuration is a chain of adjacent branes
\begin{equation}\label{eq:chain}
(0,-L)+ (0,-L+2)+\cdots + (0,L) \; 
\end{equation}
as can be deduced by the rules of Section \ref{sec:interpretation}.
In the language of Section \ref{sec:solutions} our solution has
$Q_{\alpha}=0,Q_{\tilde{3}}=P_{3}$. The decomposition of this representation 
of ${\rm su} (2)\oplus {\rm u} (1)$ gives precisely the stated result
\eqref{eq:chain}.

In the parafermion theory we have an additional $\QZ_{k}$-symmetry,
the branes $(L,0)$ and $(L,M)$ behave in the same way. Thus we can
generalize the identified processes to 
\begin{equation}\label{eq:pfprocess}
(0,M-L)+(0,M-L+2)+\cdots +(0,M+L)\longrightarrow (L,M) \ \ .
\end{equation}
We observe that all branes can be constructed out of a fundamental set
of $(0,M)$-branes.

\subsection{N=2 Minimal models}

Our results can easily be extended to the $N=2$ supersymmetric minimal 
models. The latter are obtained as
$\widehat{\rm su}(2)_k\oplus\widehat{\rm u}(1)_{2}/\widehat{\rm u}(1)_{k+2}$ 
coset theories. Now we need three integers $(l,m,s)$ to label sectors, 
where $l=0,\dots,k$, $m=-k-1,\dots,k+2$ and $s=-1,0,1,2$ are subjected 
to the selection rule $l+m+s=$ even. Maximally symmetric branes 
are labeled by triples $(L,M,S)$ from the same set. We shall restrict 
our attention to the cases with $S=0$. 

The U(1) factor in the numerator contributes an additional field X 
which enters the effective action (\ref{eq:pfaction}) minimally coupled 
to the gauge fields $\tA_a, a=1,2$. The solution (\ref{eq:pfsolution}) carries 
over to the new theory if we set X$=0$ and its interpretation is 
the same as in the parafermion case since the perturbation does not 
act in the $\widehat{\rm u} (1)_{2}$ part. It means once more that a 
chain of $P$ adjacent ($L$=0)-branes decays into a single
($L$=$P$$-$1)-brane. This process admits for  a very suggestive pictorial 
presentation. Using the geometric setting described in Section 3, we 
find the target space of the $N=2$ minimal models as a disc with $k+2$
equidistant punctures at the boundary. This was first described in
\cite{MaMoSe1}. Let us label the punctures by a $k+2$-periodic integer
$q=0,\ldots,k+1$. A brane $(L,M)$ is then represented through
a straight line stretching between the points $q_1=M-L-1$ and $q_2
=M+L+1$. In the described process, a chain of 
branes, each of minimal length, decays to a brane forming a straight 
line between the ends of the chain (see Fig.\ \ref{fig:mmbranen}). 
In \cite{HoIqVa} similar pictures occur in a geometric description
using the realization of $N=2$ minimal models as Landau-Ginzburg models.
\begin{figure}[ht]
\scalebox{.6}{\includegraphics{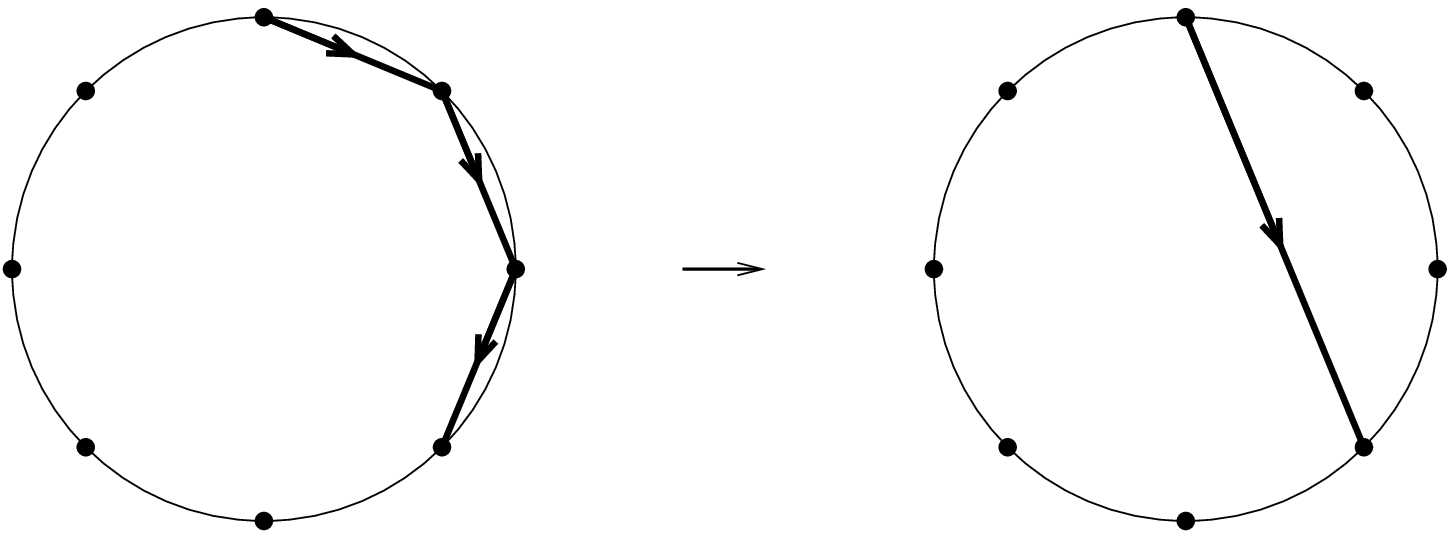}}\vspace*{-2.5cm}
\end{figure} 

\hspace*{9.95cm}\begin{minipage}{3.6cm}
{\refstepcounter{figure}\label{fig:mmbranen}Figure \ref{fig:mmbranen}:
The chain (\ref{eq:chain}) of branes can 
decay into a single brane $(L,M)$.} \bigskip \bigskip
\end{minipage}

In Figure \ref{fig:mmbranen} we have tacitly assumed that the
processes we identified in the large $k$ regime persist to finite 
values of $k$. For branes on SU(2), analogous results were described 
in \cite{AlSc2}. Similar systematic investigations in case of 
other CFT backgrounds can be performed \cite{FrSc4}. In any 
case, the results of \cite{GrRuWa1,GrRuWa2} and the comparison with 
exact studies (see e.g.\ \cite{LeSaSi}) in particular models display 
a remarkable stability of the RG flows as we move away from 
the decoupling limit. 

\subsection{Minimal models}
The minimal models are constructed as a $\widehat{\rm su}
(2)_{k}\oplus \widehat{\rm su} (2)_{1}/\widehat{\rm su} (2)_{k+1}$
coset. The embedding of the denominator theory is diagonal.
The sectors of the numerator theory are labeled by two integers
$(l,s)$ where $l=0\dots k, s=0,1$. Together with a label $l'=0\dots
k+1$ from the denominator we label the sectors of the coset model by
triples $(l,s,l')$. From the coset construction we find the selection
rule that $l+s+l'$ has to be even and the field identification
$(l,s,l')\sim (k-l,1-s,k+1-l')$. Because of the selection rule, 
$s$ is determined by fixing $l$ and $l'$ so that we can label the
sectors by pairs $(l,l')$. 

Now we want to formulate the effective action using our general formalism.
Let us again start with a configuration of branes $(L,L')$ that have trivial
label in the denominator su(2), $L'=0$. On such a configuration we have
six fields $\tA_{a},\tB_{a}$ corresponding to directions in the first 
and the second $\widehat{\rm su} (2)$-part of the numerator
respectively. The action governing the dynamics of these fields is 
constructed as in Section 4.1. The constraints \eqref{eq:genconstraint} 
translate into
\begin{equation}
\tB_{a}\ =\ -\tA_{a}-i{\tf_{ab}}^{c}\tL_{c}\tA^{b} \quad  \mbox{ for 
all } \ \ a
\end{equation}
and
\begin{equation}
\label{eq:mmconstraint2}
i\tL_{a}\tA_{b}+{\tf_{ab}}^{c}\tA_{c}\ =\ 0 \quad \mbox{ for all } 
\ \  a , c \; .
\end{equation}
By the first of these relations we can eliminate $\tB_{a}$ from the
action. The action is expanded according to powers of the level
$k$. As leading terms we find
\begin{equation}
\cS (\tA )\ =\ -\frac{1}{2k}\tr (\tL_{a}\tA_{b}\tL^{a}\tA^{b})-
\frac{2i}{3k}\tf^{abc}\tr (\tA_{a}\tA_{b}\tA_{c}) \;.
\end{equation}
Taking the $k$-dependent metric into account, we note that the action
is of order $1/k^{3}$. With the help of \eqref{eq:mmconstraint2} the 
derivatives can be eliminated and we get
\begin{equation}
\cS (\tA )\ =\ \frac{2}{k^{2}}\tr (\tA_{a}\tA^{a})-
\frac{2i}{3k}\tf^{abc}\tr (\tA_{a}\tA_{b}\tA_{c}) \;.
\end{equation}
To find solutions we have to find an extremum of this action where the
fields have to fulfill \eqref{eq:mmconstraint2}. Applying our general 
results to this example we see that we have to look for solutions
$\tA_a=Q'_a$ where the $Q'_a$ commute with the $Q_a$ and where $-Q'=S$ 
is a representation of su(2).

Let us go into an example by considering a single $(L,0)$ brane,
$L>0$. In this case the $L$+1-dimensional representation
$S_{a}=P_{a}$ is the only possibility for $S$. We can easily 
calculate the value of the action for this solution (after proper 
normalization, more details about normalization can be found in 
\cite{AlReSc2}) and obtain
\begin{equation}
\cS (-P)\ =\ \frac{\pi^{2}}{3k^{3}}L (L+2)>0 \;. 
\end{equation}
The solution describes the flow from a different brane configuration
with higher mass to the $(L,0)$-brane. From our general rules we can
identify this configuration as a single $(0,L)$-brane. 
Thus, we observe here the decay process $(0,L)\longrightarrow (L,0)$ which
coincides precisely with the results of \cite{ReRoSc}.

This process gives us the possibility to determine the effective
action of the $(0,L)$-brane by considering the fluctuation spectrum,
\begin{eqnarray}
\cS_{(0,L)} (\tA )&=&\cS_{(L,0)} (-P +\tA ) -\cS_{(L,0)} (-P )\\[2mm]
&=&+\frac{1}{2k}\tr (\tL_{a}\tA_{b}\tL^{a}\tA^{b})-
\frac{2i}{3k}\tf^{abc}\tr (\tA_{a}\tA_{b}\tA_{c}) \;,
\end{eqnarray}
which looks the same as the action for the $(L,0)$-brane except the
change of sign in front of the kinetic term. The constraint on the
fields \eqref{eq:mmconstraint2} does not change. This is the expected
result since the kinetic term comes now from the $\LA{h}$-part and
therefore comes with a different sign.

Our next example will be a configuration of one $(L,0)$-brane I and one
$(L+2,0)$ brane II. The fields $\tA_{a}$ are then described by quadratic
matrices of size $2L+4$ which we can understand as consisting of four
blocks I-I, I-II, II-I, II-II where the block I-I describes modes of
strings with both ends on brane I and so on.

\begin{equation}
\tA\ =\ \left(\parbox{3.2cm}{\makebox[3.2cm]{\myframebox{1cm}{1cm}{I-I} 
\myframebox{1cm}{1.5cm}{I-II}}
\\[2mm]  \makebox[3.2cm]{\myframebox{1.5cm}{1cm}{II-I}
\myframebox{1.5cm}{1.5cm}{II-II}}} \right) 
\parbox{2cm}{$\left.\parbox[c][1.15cm][c]{0.1cm}{} \right\} L+1 \\
\left.\parbox[c][1.65cm][c]{0.1cm}{} \right\}L+3$}
\end{equation}
The matrices $P$ which implement the derivatives can be decomposed in
$P=P^{I}+P^{II}$ where $P^{I}$ has entries only in the I-I block and
$P^{II}$ only in the II-II block.

Besides the solutions $-P^{I}$ and $-P^{II}$ we find two more
coming from a $2$-dimensional and an $L+2$-dimensional representation,
$-S^{2}$ and $-S^{L+2}$. This is easily understood because these
are just the representations appearing as tensor factors in the sum of 
representations,
\begin{equation}
[L+1]\oplus [L+3]\simeq  [2]\otimes [L+2] \; . 
\end{equation}
The value of the action is positive, so we deal here with a decay
process to the $(L,0),(L+2,0)$ system. By using the general
interpretation of Section \ref{sec:interpretation} we can 
immediately deduce the starting configuration: For the 
solution by the $2$-dimensional representation 
it is the brane $(L+1,1)$, for the other solution it is the brane
$(1,L+1)$. 

The described analysis of brane processes carries over to more general
brane configurations. We find that any $(L,L')$-brane finally decays into
a configuration with trivial denominator labels,
\begin{equation}\label{eq:mmprocess}
( L,L')\longrightarrow (\,|L-L'|,0)+ (\,|L-L'|+2,0)+\cdots + (L+L',0) \;.
\end{equation}
All branes with nontrivial label from the denominator part are
unstable and decay into configurations of branes with trivial
denominator part. Which branes appear in the decay product is
determined by the rules of how a tensor product of representations is
decomposed into irreducible representations. These are exactly the 
processes described in \cite{ReRoSc}. But our analysis shows more, 
namely that any two configurations $\sum P_{L\,L'}(L,L')$ and 
$\sum Q_{L\,L'}(L,L')$ are connected by a process if
\begin{equation*}
\sum P_{L\,L'}\, L\otimes L' \sim \sum Q_{L\,L'}\, L\otimes L' \ .
\end{equation*}
For example, any brane $(L,L')$ can be constructed as condensate 
from $L$=0-branes,
\begin{equation*}
(0,\,|L-L'|\, )+ (0,\,|L-L'|+2)+\cdots + (0,L+L')\longrightarrow
(L,L') \ .
\end{equation*}
Recently there has been a study of RG flows in minimal models \cite{Graham} 
extending the work of \cite{ReRoSc}. All fixed points discovered there by a 
thorough CFT-investigation can also be found from our general coset 
analysis.
\begin{figure}
\begin{center}
\input{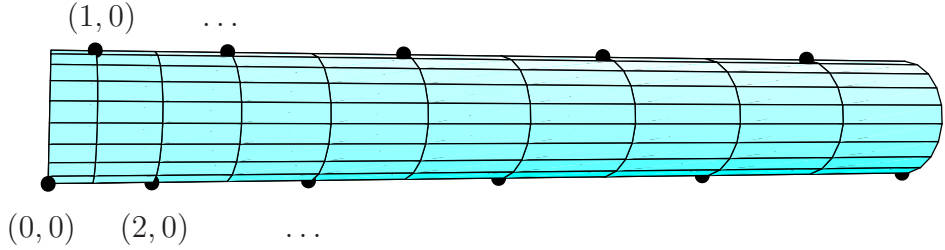}
\end{center}
\caption{\label{fig:mmodgeometry}Geometric interpretation: The 
picture shows the
underlying geometry of the minimal models together with the possible
point-like branes of the form $[L,0]$ sitting at the top and at the
bottom of a cylinder with squeezed ends. The right end of the cylinder
is cut.}
\end{figure}


\begin{figure}
\begin{center}
\input{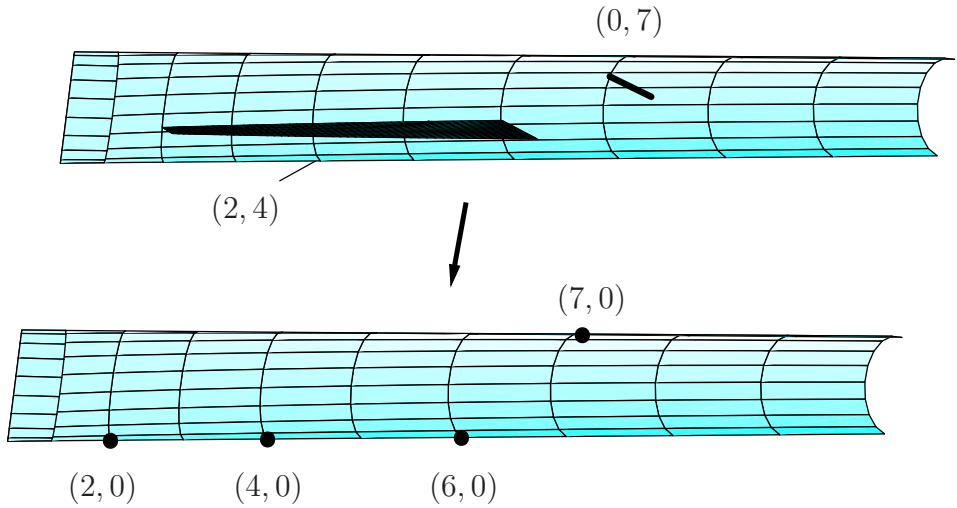}
\end{center}
\caption{\label{fig:mmodprocesses} Processes in the minimal model
geometry of Fig.~\ref{fig:mmodgeometry} with removed front wall.
Two processes are shown: 
(a) A one-dimensional string-like brane $(0,7)$
decays into one point-like brane at the top. (b) A two-dimensional
brane $(2,4)$ decays into a configuration of point-like branes at the bottom. }
\end{figure}

We can use our insights on the geometry from Section \ref{sec:bcm} 
to visualize the
processes. The target space of minimal models is a cylinder 
where the ends are squeezed to a line (see Fig.\
\ref{fig:mmodgeometry}). The simple $(L,0)$-branes are point-like branes
sitting at the top or at the bottom depending on $L$ being odd or
even. The value of $L$ varying between $0$ and $k$ determines the 
position along the cylinder (see Fig.\ \ref{fig:mmodgeometry}).
The generic branes are two-dimensional planar branes $(L,L')$. The
extension along the cylinder is between $|L-L'|$ and $L+L'$, the
vertical position is given by $L'$. If $L$ is zero these branes
degenerate to string-like branes of type $(0,L')$. Only the
point-like branes are stable, the other types of branes decay into
configurations of point-like branes as is illustrated in Fig.\
\ref{fig:mmodprocesses}. This is reminiscent of the phenomena 
that were observed in the study of tachyon condensation (see 
e.g.\ \cite{Sen1,Sen2,Sen3,ReSc2,HaKuMa}).

\section{Conclusion}
\setcounter{equation}{0} 
In this work a rather general picture of condensation processes 
in a certain limiting regime of coset models has been developed. 
We have managed to show that two brane configurations
$P$ and $Q$ on a brane are related by some flow if the restrictions 
of the corresponding representations to the diagonally embedded 
$\sh \subset \sg \oplus \sh$ are equivalent. This shows that the
conserved charges must take values in the representation ring of 
the denominator or in some quotient thereof in case there are 
further processes. Our present work can be regarded as a 
generalization of previous work in conformal field theory 
to more general brane configurations and a large class of 
coset theories. The use of non-commutative gauge theories 
made it possible to keep track of the large number of boundary 
couplings. 
\smallskip

It is of obvious interest to go beyond the limit in which some of the 
levels are sent to infinity and to study the pattern of flows for
finite values of the level, i.e.\ deep in the stringy regime. In 
case of string theory on group manifolds such an extension can be 
performed with the help of the `absorption of the boundary spin'-
principle that was formulated by Affleck and Ludwig \cite{AffLud2,
AffLud3}. We will propose an appropriate generalization of this idea 
in a forthcoming publication \cite{FrSc4}. It is interesting to 
remark that coset models typically possess brane processes at 
finite $k$ which cannot be seen in the limiting regime (see 
\cite{ReRoSc} for an example in unitary minimal models), i.e.\ 
these condensation processes are not deformations of a process
one can study in the `geometric regime'. 
\smallskip

An obvious extension of our analysis is to go beyond the Cardy 
case and to incorporate e.g.\ boundary theories that are obtained 
from branes localized along the twined conjugacy classes on 
the group manifold $G$ \cite{FFFS1}. The latter arise when we 
glue left- and right moving currents of the WZW-model for the 
group $G$ with some automorphism $\Omega$ of $\sg$, 
\begin{equation} J^\a(z) \ = \ \Omega (\bar J^\a)(\bz) \ \ 
\mbox{ for } \ \ \a = 1,\dots, D = \dim G \ \ . 
\label{eq:glue1} \end{equation}
The associated branes have been shown to be localized along 
the following twined conjugacy classes 
$$ C^{G; \omega} _g \ = \ \{ \, g' \in G \ | \ g' = u g \omega(u^{-1}) 
\mbox{ for } u \in G \ \} \ \ . $$
Here, $\omega$ denotes the automorphism of the group $G$ that 
comes with $\Omega$. These twined branes on group manifolds 
descend to the coset $G/H$ provided that $\omega$ can be 
restricted to the subgroup $H \subset G$. In the latter case 
it induces an automorphism on $H$ and we can construct the 
corresponding twined conjugacy classes $C^{H;\omega}_h$. The 
induced branes of the coset model are  localized along 
$$  C^{G/H;\omega}_{(g,h)} \ = \ \pi^{G}_{G/H}\left(\, 
    C^{G;\omega}_g \ (C^{H;\omega}_{h})^{-1} \, 
    \right) \ \subset \ G/H \ \ . 
$$       
To show that this prescription is consistent one has to show
that the adjoint action of $H$ on $G$ leaves the space 
$C^{G;\omega}_g \ (C^{H;\omega}_{h})^{-1}$ invariant. The 
dynamics of such twined branes in coset models can be studied
once more by a reduction from the theory of twined branes on 
group manifolds. The latter was constructed recently in 
\cite{AFQS}. 
\bigskip  

\noindent
{\bf Acknowledgments:} We would like to thank I.\ Brunner, 
M.\ Gaberdiel, K.\ Gawedzki, K.\ Graham, T.\ Quella, 
A.\ Recknagel and N.\ Reis for stimulating and very useful 
discussions.


\begin{appendix}
\section{On brane geometry in coset models}
\setcounter{equation}{0} 
In this appendix we plan to derive the geometric interpretation 
(\ref{cosgeom}) of D-branes in coset models that was found in 
\cite{Gaw2} from the 1-point functions (\ref{cos1pf}). The main 
idea is borrowed from an analogous discussion of brane 
geometry on group manifolds \cite{FFFS1} and it generalizes 
the constructions of \cite{MaMoSe1}. 
\medskip

It is useful to recall very briefly how one decodes the brane 
geometry on group manifolds (\ref{ggeom}) for the 1-point 
functions (\ref{eq:CSCg}). To begin with one needs to set up  
a correspondence between functions on the group and the bulk
fields whose conformal dimension vanishes when we send
$k$ to infinity, i.e.\ the fields listed in (\ref{eq:list1}). 
This correspondence is obvious since the space $\Fun(G)$ of 
functions on the group $G$ is spanned by the matrix elements 
$D^\la_{nm}(g)$ of irreducible representations where $\la$ 
runs through the set $\cJ^\sg$. With such a relation between 
bulk fields and functions in mind, the formula (\ref{eq:CSCg}) 
suggests to introduce a set of functions $T^\sg_\La: G 
\rightarrow \QC$,   
$$ T^\sg_\La(g)  \  := \ \sum_{\la \in \cJ_k} \    
   \frac{S^\sg_{\La \, \la}}{\sqrt{S^\sg_{0\, \la}}} \ 
   \delta_{nm} \ D^\la_{nm}(g) $$ 
in which the basis element $D^\la_{nm} \in \Fun(G)$ is 
weighted with the strength of the coupling of the associated 
closed string mode to the brane $\La$. The function $T^\sg_\La$ 
can be shown to possess a peak along a conjugacy class 
$ C^G_\La$ of $G$ \cite{FFFS1}. This confirms the geometrical 
interpretation of the gluing condition (\ref{eq:glue}) uncovered 
in \cite{AlSc1}. 
\medskip 

After this preparation we want to turn to the case of branes in 
a coset $G/H$. Now we need to find a correspondence between bulk 
fields from the set $\cJ^r$ and a set of functions on  
the coset space $G/H$ where the denominator $H$ acts on $G$ by 
conjugation. To construct such functions, we rewrite $G/H$ as 
a coset of the form $G \times H/ H\times H $ in which the two 
factors $H$ in the denominator act by left and right multiplication 
on $G \times H$, respectively,   
$$ u_l (g,h) \ = \ (u g, u h) \ \ \ , \ \ \ 
   v_r (g,h) \ = \ (g v^{-1}, h v^{-1} ) $$ 
for all $u,v \in H$ and $(g,h) \in G \times H$. The equivalence of 
the two coset constructions is based on the equality $G \times H / H 
= G$ and it uses the fact that after dividing out one copy of $H$ 
from $G \times H$, the second factor $H$ acts by conjugation on $G$ 
rather than by left- or right translation.   
\smallskip

Our aim now is to argue that there exists a correspondence between the coset 
fields labeled by $\cJ^r$  and functions on $G \times H / H \times H$, i.e.\  
$H \times H$-invariant functions on $G \times H$. To this end, let us 
note that such invariant functions are obtained by averaging elements 
of $\Fun(G \times H)$ over the group $H \times H$ of translations, i.e.\ 
$$ \int_{H\times H} d\mu_{H\times H} (u,v) \ F(u g v^{-1}, 
u h v^{-1})  \ \ . $$ 
If we apply this averaging prescription to the basis $D^\la_{nm}(g) 
\bD^{\la'}_{rs}(h)$ of $\Fun(G \times H)$ we obtain a non-vanishing 
invariant function whenever the representation $\la'$ of the finite 
dimensional Lie algebra $\sh$ is contained in $\la$. This means that 
for each bulk field labeled by elements from the set $\cJ^r$ there 
exists a function on $G \times H/ H \times H$. We can now apply the 
same procedure as in the WZW case to read off the geometry of the 
branes in the coset theory, i.e.\ we define a function 
\begin{eqnarray}  T_{(\La,\La')} & = & \sum_{(\la,\la')} 
   \frac{S^\sg_{\la\,\La} \bar S^{\sh}_{\la'\,\La'}}{\sqrt{ 
   S^\sg_{0\, \la} \bar S^{\sh}_{0\la'}}}
   \int d\mu_{H\times H}(u,v) 
   D^\la_{mm} (ugv^{-1}) \bD^{\la'}_{ss} (u h v^{-1})
   \nn  \\[2mm]
   & = & 
   \int d\mu_{H\times H}(u,v) 
   T^\sg_\La (ugv^{-1}) \overline{T}^\sh_{\La'} 
   (u h v^{-1}) \ \ . 
\end{eqnarray}
Since $T^\sg_\La(g) \overline T^\sh_{\La'} (h)$ is localized
along the product $C^G_\La \times C^H_{\La'} \subset G 
\times H$, we have just shown that the coset brane is localized 
along the image of this product in the coset space $ G \times H 
/ H \times H$. Rephrased in terms of the more conventional coset 
$G/H$ this means that the coset brane $(L,L')$ is localized along 
the image of the space $C^G_\La (C^H_{\La'})^{-1}$, 
in agreement with eq.\ (\ref{cosgeom}).

\section{Effective action for general branes}
\def\bM{\bar M}
\def\bN{\bar N} 
\def\bn{\bar n} 
\setcounter{equation}{0} 
It this section we sketch the derivation of the effective action 
for a general coset brane configuration from conformal field 
theory. We begin by looking at the following product $\H^{\sg\ M}_L
\otimes \H^{\sh\ \bM'}_{\bL'}$ of state spaces for boundary theories of 
the $G$ and $H$ WZW model. Within such a space we want to find the 
state space $\H_{(L,L')}^{\ (M,M')}$ of the coset theory. In a first 
step let us impose the constraints 
\begin{equation}\label{const1} 
J^{i}_{n}\ \psi \ = \ J^{\im}_{n}\ \psi \ = \ 0 \quad 
 \ \mbox{ for all } \  n>0 \; , \;
\end{equation}
and $i,\im$ run through the usual range. This restricts us to 
the ground states for the actions of $\hh \subset \hg$ on the 
first factor and of $\hh$ on the second. With the help of eq.\
(\ref{eq:cosOSdec}) we can conclude that the resulting subspace 
of states satisfying eqs.\ (\ref{const1}) has the form  
\begin{equation}
\bigoplus_{l,m,n} \ N^{\sg ; M}_{L\, l} \, \cH_{(l,m)} \otimes V_{m}^{\sh}
  \otimes  N^{\sh ; \bM'}_{\bL'\, n}  V_n^\sh \; 
\end{equation}
where $V_m^\sh$ denotes the space of ground states  in $\H^\sh_m$ and
we sum over all $m$ such that $(l,m)$ is a sector of the coset model. 
If we now require the additional invariance condition  
\begin{equation} \label{const2} 
(J^{i}_{0}+\tilde{J}^{i}_{0}) \ \psi \ = \ 0 
\end{equation} 
then the only contribution in the sum will come from $m=\bn$ and the 
invariant part of $V_{m}^{\sh}\otimes V_{n}^{\sh}$ is one-dimensional. 
This means that after imposing the two constraints (\ref{const1},\ref
{const2}), we are left with the space 
\begin{equation}
\bigoplus_{l,m} \ N^{\sg ; M}_{L\, l} \,  N^{\sh ; M'}_{L'\, m} 
 \ \cH_{(l,m)} 
\end{equation}
which is isomorphic to the state space $\H_{(L,L')}^{(M,M')}$ 
of the boundary coset model. In this way we have prepared states
of the coset theory from states of the product of boundary WZW 
models. 
\medskip 

Now we use the boundary operators of the WZW models to build
boundary operators on the product space. These will then be 
shown to reduce to boundary fields of the coset theory when 
$k$ is sent to infinity. The idea is to use fields of the 
form 
\begin{equation} 
J^a\,  \Psi^{N L}_{(l,\nu)}\,  \Psi^{\bN' \bL'}_{(l',\nu')}\,  
    C_{a; \nu, \nu'}^{ll'} \ : \H_{L}^{\sg\, M} \otimes 
    \H_{\bL'}^{\sh\, \bM'} \ \rightarrow\  \H_{N}^{\sg\, M} 
    \otimes \H_{\bN'}^{\sh \, \bM'}  
\label{op} 
\end{equation} 
where $\nu, \nu'$ label a basis in the representation spaces 
$V^{l}, V^{l'}$, respectively, and the coefficients $C$ are 
chosen such that the operator is invariant under the obvious 
action of $\sh$. This choice of $C$ guarantees  that the 
operators respect the constraint (\ref{const2}). On the other 
hand, they are 
not compatible with our first set of constraints (\ref{const1}) 
simply because boundary primary fields usually map ground 
states into a linear combination which contains also excited 
states. But these excitations get suppressed for large values 
of the level so that the operators (\ref{op}) do respect the 
conditions (\ref{const1}) in the limiting regime and hence 
they become the operators of the boundary theory 
at $k \rightarrow \infty$. This means that we have reduced 
the computation of $3^{rd}$ and $4^{th}$ order terms in our 
effective field theory to computations in the boundary WZW 
model for $G$ and $H$. These calculations have been performed in 
\cite{AlReSc2} and they provide the corresponding terms in 
the action (\ref{pWZWact}). But in the case of coset theories,
we work only with a small subset of boundary fields from the 
WZW models which is specified by the constraints (\ref{const1},
\ref{const2}). They manifest themselves in the constraints 
(\ref{eq:genconstraint}) of the effective field theory. 
\smallskip

It remains to discuss the quadratic terms in our effective action. 
These terms can be read off from the conformal dimensions. More 
precisely a mode $(l,l')$ of the coset model contributes a 
quadratic term proportional to $h_{(l,l')}$. 
But in our construction of the theory from the two WZW models, 
$(l,l')$ is accompanied by the field of weight $h_{l'}$ for the 
subalgebra $\hh \subset \hg$ and another field with the same 
weight being associated with the second WZW model. This would 
add up to $h_{(l,l')} + 2 h_{l'} \neq h_{(l,l')}$. Our prescription
to put an extra factor $\sqrt{-1}$ into the derivatives $\tL_{\im}$, 
accounts for the mismatch. This is due to the fact that the conformal 
weights are obtained from the quadratic Casimir which changes
sign under the replacement $J \rightarrow iJ$. Hence, the extra
factor $i$ does produce the right quadratic terms $h_{(l,l')} + 
h_{l'} - h_{l'}= h_{(l,l')}$ in the effective action. It is easy 
to see that it does not change the higher order terms in the 
constrained model. 

\end{appendix}
 
\def\gaw{Gawedzki}

\end{document}